\documentclass[runningheads]{llncs}
\makeatletter
\renewcommand\subsubsection{\@startsection{subsubsection}{3}{\z@}%
                       {-18\p@ \@plus -4\p@ \@minus -4\p@}%
                       {0.5em \@plus 0.22em \@minus 0.1em}%
                       {\normalfont\normalsize\bfseries\boldmath}}
\makeatother
\usepackage{makecell}
\usepackage{graphicx}
\usepackage{cite}
\usepackage{amsmath,amssymb,amsfonts}
\usepackage{algorithmic}
\usepackage{graphicx}
\usepackage{textcomp}
\usepackage{xcolor}
\usepackage{comment}
\usepackage{hyphenat}
\usepackage{multirow}
\usepackage{tabu}
\usepackage{caption}
\usepackage{subcaption}
\usepackage{
tikz,
relsize,
amsmath,
booktabs,
tikz
}
\usetikzlibrary{arrows,calc,positioning,shadows,shapes}
\usepackage{pgfplots}
\usepackage{pgfplotstable}
\pgfplotsset{compat=1.6}

\usepackage[T1]{fontenc}
\usepackage[utf8]{inputenc}
\usepackage{lmodern}

\usepackage[
labelfont=sf,
hypcap=false,
format=hang,
width=0.8\columnwidth
]{caption}

\usetikzlibrary{
calc,trees,shadows,positioning,arrows,chains,
decorations.pathreplacing,
decorations.pathmorphing,
decorations.shapes,
decorations.text,
shapes,
shapes.geometric,
shapes.symbols,
matrix,
patterns,
intersections,
fit}

\pgfdeclarelayer{background layer}
\pgfdeclarelayer{foreground layer}
\pgfsetlayers{background layer,main,foreground layer}

\tikzset{
>=latex
}
\usepackage{amsmath}
\usepackage{amssymb}
\usepackage{arydshln}
\usepackage{mathrsfs}
\usepackage[export]{adjustbox}
\usepackage{float}

\usepackage{times}
\usepackage{fancyhdr,graphicx,amsmath,amssymb}
\usepackage[ruled,vlined]{algorithm2e}

\def\BibTeX{{\rm B\kern-.05em{\sc i\kern-.025em b}\kern-.08em
    T\kern-.1667em\lower.7ex\hbox{E}\kern-.125emX}}

\usepackage[T1]{fontenc}
\usepackage{graphicx}
\usepackage{tikz}

\begin{document}
\include{pythonlisting}
%
\title{DCC: A Cascade based Approach to Detect Communities in Social Networks}

%
%
\author{Soumita Das \inst{1}\orcidID{0000-0002-2412-9525} \and
Anupam Biswas \inst{1}\orcidID{0000-0003-0756-6026}\and
Akrati Saxena \inst{2}\orcidID{0000-0002-7151-6309}}
%

%

\institute{Department of Computer Science and Engineering, \\National Institute of Technology Silchar, Silchar, Assam, India \\ 
\and Eindhoven University of Technology, The Netherlands \\
\email{\{soumita\_rs,anupam\}@cse.nits.ac.in,a.saxena@tue.nl}}

\maketitle              
\begin{abstract}
Community detection in Social Networks is associated with finding and grouping the most similar nodes inherent in the network. These similar nodes are identified  by computing tie strength. Stronger ties indicates higher proximity shared by connected node pairs. This work is motivated by  Granovetter's argument that suggests that strong ties lies within densely connected nodes and the theory that community cores in real-world networks are densely connected. In this paper, we have introduced a novel method called \emph{Disjoint Community detection using Cascades (DCC)} which demonstrates the effectiveness of a new local density based tie strength measure on detecting communities. Here, tie strength is utilized to decide the paths followed for propagating information. The idea is to crawl through the tuple information of cascades towards the community core guided by increasing tie strength. Considering the cascade generation step, a novel preferential membership method has been developed to assign community labels to unassigned nodes. The efficacy of $DCC$  has been analyzed based on quality and accuracy on several real-world  datasets and baseline community detection algorithms.

\end{abstract}
\section{Introduction}

Online Social Networks (OSNs) consists of inherent modular structures called communities where, nodes within a community are densely connected, and, nodes between communities are sparsely connected. Moreover, OSNs is predominantly used for information sharing because of it's ability to connect geographically distant users. As information sharing occurs through social contacts, so the underlying network structure plays an important role in information propagation. Studying and analyzing the connections of the underlying network structure is vital for solving the problem of information diffusion and hence, community detection. In OSNs, the strength of the connections shared by users are different. Numerous local similarity measures have been proposed to compute the strength of these connections using local neighborhood similarity, such as, Jaccard Index (JI), Preferential Attachment (PA), Salton Index (SA), etc. These local similarity measures are particularly beneficial in community detection because it has low time complexity. For e.g.  $(\alpha, \beta)$ algorithm utilizes JI to identify communities. 
\vspace{10pt}
\begin{algorithm}[t]
      \scriptsize \caption{Disjoint Community detection using Cascades}
        \SetKwInput{KwInput}{Input}                
        \SetKwInput{KwOutput}{Output}             
        \DontPrintSemicolon
 
        \KwInput{\indent \ \textit{G(V,E)}, \textit{p},  \textit{$U$}}
        \KwOutput{\textit{$C=\{c_{0},c_{1},..,c_{k}\}$}: set of communities}

     \SetKwFunction{ROSE}{DCC}
     \SetKwFunction{shares  Information at Sensors}{SIS}

     \SetKwProg{Fn}{Procedure}{:}{\KwRet}
     \Fn{\ROSE{$G(V,E), p, U$}}
     {
        \scriptsize $A$ \ \ $\leftarrow$ \scriptsize empty list\\
        \scriptsize $U$ $\leftarrow$ \scriptsize list of all nodes\\
        \scriptsize $C_{l}$ $\leftarrow$ \scriptsize empty list :~{stores lists of cascades}\\
        \scriptsize $C$ $\leftarrow$ \scriptsize empty list :~{stores lists of communities}\\
          \scriptsize $p$ $\leftarrow$ \scriptsize select any random node from $U$\\
          remove(p,U)\\
          add(p,A) \\
          \scriptsize $path\_length=1$\\
          \scriptsize $q$ $\leftarrow$ \scriptsize find\_maxts(p, $\Gamma(p)$)\\
          
          \scriptsize \While {len($U$)>0}
          {
          \scriptsize $r$ $\leftarrow$ find\_maxts(q,$\Gamma(q)$)\\
          \scriptsize \If{$NS(p,q)$ $<=$ $NS(q,r)$}
             {
                 \scriptsize store $p$ in $A$ \\
                 \scriptsize remove $(q,U)$ \\
                 \scriptsize $path\_length++$ \\
                 \scriptsize $t$ $\leftarrow$ $q$ \\
                 \scriptsize $q$ $\leftarrow$ $r$ \\
              }         
         \scriptsize \ElseIf {$path\_length > 1$}
            {
                  \scriptsize store $q$ in $A$ \\
                  \scriptsize store $A$ in $C_{l}$\\
                  \scriptsize make  empty $A$\\
                  \scriptsize $t$ $\leftarrow$ $r$\\
            }          
      \scriptsize \Else
            {
                 \scriptsize $t$ $\leftarrow$ $q$ :{start new process with $q$}\\
         
            }  
            \scriptsize \If{$t$ in $A$ and $U$ not empty} 
            {
            $p$ $\leftarrow$ select any random node from $U$
            }
            
        }
        \scriptsize Assign community labels to all respective cascades in $C_{l}$ and store in $C$ 
      
        \scriptsize \While {U not empty}
         { 
            \scriptsize  $PM(u,c_{j})$ computed using equation
                 $\forall$ $u$ $\in$ $U$, $\forall$ $c_{j}$ $\in$ $C$

             $C$ updated with addition of unassigned nodes to respective communities
        }
    
        \scriptsize \If{ ~two~ communities~ say, $c_{1}$ $\in$ $C$, $c_{2} \in C $ share atleast 1 node}
        {Merge($c_{1}$,$c_{2}$)}      
        \scriptsize  \KwRet $C$  
        }       
\end{algorithm}

Social network analysis is predominantly associated with analyzing the interaction patterns among people, states or organizations. These interactions among users helps to reveal various important details of the underlying network structure\cite{das2021deployment}. The interactions in OSNs is dependent on the relationships shared by the connected users. These relationships are analyzed using tie strength measure. Strong ties cover densely knitted networks~\cite{strgr} and this idea is used to design a novel tie strength measure which is contingent on the neighborhood density of connected node pairs. Next, the tie strength is utilized to guide the interactions among individuals. Basically, $DCC$ utilizes the interaction paths to reach the core of communities.  Studies suggest that community cores are most densely connected~\cite{commcore}; so, if we start the diffusion process from any node and approach towards the community core, the tie strength goes on increasing.  Tracing all the interaction patterns is used for ensembling groups of similar nodes. Therefore, our work shows the effectiveness of the proposed tie strength measure and information diffusion strategy on the identification of optimal communities. In this paper, the primary contribution is the introduction of a cascade based method called Disjoint Community detection using Cascades ($DCC$) which shows the significance of tie strength, neighbors of neighbors and information diffusion for detection of communities.

\begin{algorithm}[t]
        \scriptsize \caption{Neighborhood Similarity}
        \SetKwInput{KwInput}{Input}                
        \SetKwInput{KwOutput}{Output}             
        \DontPrintSemicolon
 
        \KwInput{\indent \ \textit{G(V,E)}, \textit{p}, \textit{q}}
         \KwOutput {\textit{$ts$}: Tie strength value}

     \SetKwFunction{ROSE}{NS(p,q)}
     \SetKwFunction{Find suitable snowball radius }{FSSR}

     \SetKwProg{Fn}{Procedure}{:}{\KwRet}
     \Fn{\ROSE}
     {
     \scriptsize \If{$\Gamma(p)==1$ or $\Gamma(q)==1$}
     {
     \scriptsize $ts=0$\\
     
     }
     \Else{
     \scriptsize $ts=\frac{\rho_{p,q}}{\mid \chi_{pq}\mid}$\\
     }
     \scriptsize \KwRet $ts$ \\
     }
\end{algorithm}

\begin{algorithm}
\caption{Find Maximum Tie Strength}
\SetKwInput{KwInput}{Input}                
\SetKwInput{KwOutput}{Output}             
\DontPrintSemicolon
 
\KwInput{\ $G(V,E), p, \Gamma(p)$ \\}
\KwOutput{\textit{$q$}: neighbor of $p$ sharing maximum tie strength with $p$}

\SetKwFunction{TSE}{find\_maxts}

\SetKwProg{Fn}{Procedure}{:}{\KwRet}
\Fn{\TSE{$G(V,E), p, \Gamma(p)$}}
{
   \scriptsize $max=0$ \\
   \scriptsize \For{$q$ in $\Gamma(p)$}
   {
   \scriptsize \If {$NS(p,q)$ > $max$}
   {
   \scriptsize $max=NS(p,q)$
   
   }
   }
\scriptsize \KwRet $q$ \\
     }
\end{algorithm}

The rest of the paper is organized as follows. Section~\ref{relatedwork} discusses about the related work, Section~\ref{pro} briefs about the proposed cascade based community detection method, Section~\ref{four} discusses about the experimental setup, Section~\ref{five} discusses about the result analysis and Section~\ref{six} concludes the paper. 

\section{Related Work}
\label{relatedwork}
Increase in the size of social media users has made social network analysis very complex. Therefore, community detection task has been introduced to reduce the complexity of the original network in a substantial manner. Moreover, there are several potential applications of communities in OSNs such as, it is used in recommendation systems, trend analysis in citation networks, evolution of communities in social media, discovering fraudulent telecommunication network activities, dimensionality reduction in pattern recognition. Therefore, several community detection techniques have been introduced till date to identify communities which are primarily classified into several approaches based on graph partitioning, clustering, modularity optimization, random walk and diffusion community~\cite{chand2017community,khan2017network}. Spectral Bisection method is a graph partitioning technique which divides the graph into clusters based on density of links within a cluster and between clusters~\cite{pothen1997graph,barnes1982algorithm}, $Gdmp2$~\cite{chen2010dense} is a clustering technique where set of similar nodes are grouped together. It is usually of two kinds such as, hierarchical clustering~\cite{hastie2009elements} and partitioning method of clustering~\cite{hlaoui2004direct,bezdek2013pattern}, Greedy-modularity $(GM)$~\cite{clauset2004finding}, $Kcut$~\cite{ruan2007efficient} are modularity maximization techniques which are based on partitioning the graph based on the best modularity value~\cite{newman2004finding,kirkpatrick1983optimization,boettcher2002optimization,newman2006modularity,holland1992adaptation,jin2021survey}, Diffusion Entropy Reducer $(DER)$~\cite{kozdoba2015community} uses random walk technique where communities are detected by adopting a walker where the overall time is dependent on the density of communities~\cite{hughes1995random,zhou2003distance},  Label Propagation Algorithm $(LPA)$~\cite{cordasco2010community} utilizes diffusion community method where similar nodes are grouped by propagating same action, property or information in a network. 

\section{Proposed Method}
\label{pro}


The social contacts shared by an individual is indicative of some similarity possessed by the corresponding individuals, but mere connection is not enough to determine the most similar nodes present in the network. $DCC$ addresses the role of tie strength and cascades in the identification of communities inherent in a network. In this section, we shall discuss the preliminary concepts that would be used throughout this paper followed by the discussion of $DCC$ algorithm in detail.

\subsection{Preliminaries}
\label{pre}
Suppose, we consider a graph $G(V, E)$ where $V$ refers to set  of  nodes, $E$ refers to set of edges. For any node $v \in V$, set of neighbors of $v$ is denoted by $\Gamma(v)$, degree of node $v$ is indicated by $\mid \Gamma(v) \mid$. Then, for a connected node pair $(v,u) \in V$, number of connections shared by common neighbors of $v$ and $u$ is indicated by $\mid \sigma(vu) \mid$.

\begin{definition}(Unprocessed Node). Given a graph $G(V,E)$, a node $p \in V$ is an unprocessed node, if $p$ is not yet activated during the diffusion process.
\end{definition}

\begin{definition}(Common Neighborhood). Given a graph $G(V,E)$ and a connected node pair say, $e_{p,q} \in E$, then Common Neighborhood is used to find the neighboring nodes related to $p$ and $q$. Common neighborhood of $(p,q)$ pair is defined by,


\begin{align}
 \label{cn}
 \scriptsize
   \rho_{p,q} =&|\Gamma(p) \cap \Gamma(q)|+ |\Gamma(p) \cap \Gamma(z)| + \left|\Gamma(q) \cap \Gamma(z)| +
  |\sigma_{pq}| \right.\nonumber\\ &|\Gamma(w)  \cap \Gamma(z)|, \forall (w,z) \in \Gamma(p) \cap \Gamma(q) \nonumber\\  &~if~  e_{w,z} \in E,  ~w \ne z, 
\end{align}
\end{definition}
which indicates that $NS(p,q)$ is dependent on the degree exhibited by $p$ and $q$. If either of the node's degree is 1, then $NS(p,q)=0$.

\begin{definition}(Neighborhood Similarity).  Given a graph $G(V,E)$, Neighborhood Similarity of an edge say, $e_{p,q} \in E$ indicates the tie strength of $p$ and $q$. It is defined by,   

\begin{equation}
\label{ns}
NS_{p,q}=\frac{\rho_{p,q}}{\mid \chi_{p,q}\mid,}
\end{equation}
\end{definition}
where, numerator term refers to common neighborhood of $(p,q)$ pair and denominator term indicates number of nodes belonging to common neighborhood of $p$ and $q$.

\begin{definition}(Cascade). A cascade is a tuple $(p,find\_maxts(p,\Gamma(p), T)$ which contains information about a node $p$, neighbor of $p$ with which $p$ shares maximum $NS$ score indicated by $find\_maxts(p,\Gamma(p))$ at a certain time $T$.
 \end{definition}
 
It is important to understand how cascades are generated during the diffusion process and how these are used for identifying communities present in the network. Therefore, it is required to understand the $DCC$ algorithm to obtain a concrete idea of the community detection process. The details of the $DCC$ algorithm is discussed below. 

\begin{figure}[ht]
\centering {\includegraphics[width =9.5cm]{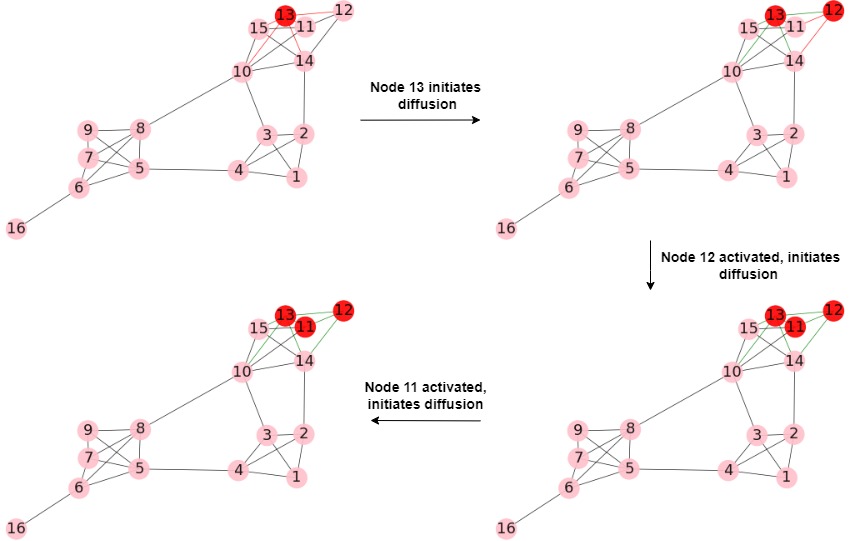}}
 \captionsetup{width=\columnwidth}
     \caption{\scriptsize Demonstration of cascade generation step of $DCC$ algorithm using a simple graph. Peach colored nodes indicate inactive nodes, red colored nodes indicate active nodes, red arcs represent edges where active nodes try to activate their inactive neighbors, green arcs represent edges propagated once. Initially, node 13 initiates diffusion process, computation of $Find\_maxts(13,\Gamma(13))$=12 with $NS(13,12)=0.8$, so node 12 activated and initiates diffusion, computation of  $Find\_maxts(12,\Gamma(12))$=11, $NS(12,11)=1.0$ and $NS(13,12)<=NS(12,11)$, node 11 is activated and initiates diffusion, computation of  $Find\_maxts(11,\Gamma(11))$  gives no neighbor of node 11 that shares greater tie strength than $NS(12,11)$, cascade obtained is [13,12,11].}
      \label{demonstration}
\end{figure}

 
\subsection{Disjoint Community detection
using Cascades} It is a cascade based disjoint community detection approach. $DCC$ comprises of three steps. Firstly, cascades are generated by computing and comparing tie strength based on Neighborhood Similarity measure. Secondly, Preferential Membership method is proposed to assign community labels to the unprocessed nodes and thirdly, merging step where communities sharing common nodes are merged.  Let us now try to understand each of the steps with the help of pseudocodes and pictorial example.

\textbf{Cascade Generation:}  The path followed during  information diffusion process is dependent on a novel tie strength measure called Neighborhood similarity ($NS$). Tracing the path generated during the diffusion process results in a set of cascades as shown in Fig.~\ref{demonstration}. Let us try to understand the cascade generation step illustrated in the first while loop in Algorithm 1 with the help of the cascade generation example on the simple graph as shown in Fig.~\ref{demonstration}. Suppose, node 13 (indicated by red colored node) initiates the diffusion process. Then, node 13 tries to activate it's maximum $NS$ value neighboring node obtained using $Find\_maxts$. The illustration of $Find\_maxts$ is shown in Algorithm 3. Computation of $Find\_maxts(13,\Gamma(13))$ gives node 12 with $NS(13,12)=0.8$. Next, node $12$ is activated. Next, node 12 tries to activate it's neighboring nodes indicated with red arcs. The task is to identify the neighboring node say $r$ such that, $Find\_maxts(12,\Gamma(12))=r$~(say) and $NS(13,12)<=NS(12,r)$. We find $r$=node 11 with $NS(12,11)=1.0$ and hence, node 11 is activated, indicated with red color. Now, node 11 tries to find it's maximum $NS$ value neighbor such that it's tie strength is greater than or equal to $NS(12,11)$. But, no such suitable neighboring node is obtained and hence, the cascade obtained is $[13,12,11]$. Next, all cascades for the remaining unprocessed nodes are obtained by repeating the above mentioned procedure.  At the end of the cascade generation step, a list of cascades are obtained which are assigned with corresponding community labels. Next, labels are assigned to remaining unlabelled nodes using Preferential Membership ($PM$).       \\

\begin{definition}(Preferential Membership). Given graph $G(V,E)$, set of communities $C$; then, Preferential Membership is used to assign community membership $c_{j} \in C$ to an unlabelled node, $p \in V$ when $\underset{j}{\operatorname*{arg\,max}} ~~PM(p,c_{j}),~ \forall c_{j}\in C$. It is defined by,

\begin{align}
   \label{pm}
    PM(p,c_{j})= \sum_{\substack{\Gamma(q) \neq p,\\ q \in \Gamma(p),\\q \in c_{j}}}\frac{\mid \Gamma(p) \cap \Gamma(q)\mid}{\mid \Gamma(p) \mid  \times \mid \Gamma(q) \mid }
   \end{align}
\end{definition}

Nodes that are yet to be labelled are assigned with corresponding community labels using equation~\ref{pm}. Let us try to understand the membership assignment with an example. Suppose, we assign a community label $c_{1}$ to the  cascade [13,12,11] as obtained from the cascade generation process. Now, node 10 (say) is one of the unprocessed node, then using equation~\ref{pm}, we compute $PM(10,c_{1})$.  Considering this equation, we select node 13 which is one of the neighbors of node 10. Moreover, node 13 is also in $c_{1}$. Neighbors of node 13 is indicated by, 

$\begin{array}{lcl}
\Gamma(13)&=& \{10,12,11,14,15\}.\\
\end{array}$\\
Next, to compute equation~\ref{pm}, we need,

$\begin{array}{lcl}
 \Gamma(12) & = &\{11,13,14\}.\\
 \Gamma(11)& = & \{10,12,13,15\}.\\
 \Gamma(14)& = & \{2,10,12,13,15\}.\\
 \Gamma(15)& = & \{10,11,13,14\}.\\
\end{array}$\\
Now, putting these values in equation~\ref{pm}, we obtain,\vspace{0.1cm}

$\begin{array}{lcl}
  PM(10,c_{1}) & = & \frac{\mid \Gamma(10) \cap \Gamma(12)\mid + \mid \Gamma(10) \cap \Gamma(11) \mid+ \mid \Gamma(10) \cap \Gamma(14)\mid + \mid \Gamma(10) \cap \Gamma(15) \mid }{\Gamma(10) \times \Gamma(13)} \\
  & = & \frac{3+2+2+3}{6 \times 5} \\
  PM(10,c_{1}) & = & 0.333.
\end{array}$\\

\begin{figure}[t]
\centering {\includegraphics[width =4.2cm]{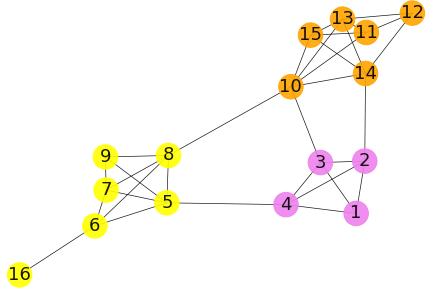}}
\caption{\scriptsize Communities obtained by $DCC$ algorithm on karate dataset. Three different colors indicate three different communities obtained by incorporation of $DCC$ algorithm.}
\label{fc}
\end{figure}

\textbf{Merging:} Merging is incorporated to obtain the final community set. Communities sharing at least one common node are merged. The final set of communities obtained by $DCC$ algorithm on the example graph is shown in Fig.~\ref{fc}. Therefore, incorporation of $DCC$ algorithm gives three set of communities. 

\begin{table}[t]
\caption {\scriptsize Dataset Statistics. First column contains dataset details, $\# ~Nodes$ refers to number of nodes, $\# ~Edges$ refers to number of edges, Avg. degree indicates average degree of the graph.}

\label{dat}
\centering
 \begin{tabular} {|m{5.2em}|m{3em}|m{3em}|m{3em}|m{5.2em}|m{3em}|m{3em}|m{3em}|}
 \hline
 \centering{Dataset} & \centering{\# ~Nodes} & \centering{ \# ~Edges} & \centering{Avg. degree} & \centering{Dataset} & \centering{\# ~Nodes} & \centering{ \# ~Edges} & \centering{Avg. degree} 
 \tabularnewline [4pt]
\hline
  \centering{Riskmap~\cite{riskmap}} & \centering{42} &  \centering{83}  &  \centering{3.95} & \centering{Dolphin~\cite{dolphin}} & \centering{62} &  \centering{159}  &  \centering{5.12} 
  \tabularnewline[4pt]
  \centering{Karate~\cite{karate}} & \centering{34} &  \centering{78}  &  \centering{4.58} & \centering{Strike~\cite{strike}} & \centering{24} &  \centering{34}&  \centering{3.16}
  \tabularnewline[4pt]
  \centering{Football~\cite{football}} & \centering{115} &  \centering{613}&  \centering{10.66} & \centering{Sawmill~\cite{sawmill}} & \centering{36} &  \centering{37}  &  \centering{3.44} 
    \tabularnewline [4pt]
   \hline
   \end{tabular}
\end{table}

\pgfplotstableread[row sep=\\,col sep=&]{
NGM & DCC & LPA & GM & DER & Gdmp2 & Kcut\\
Riskmap & 0.6293  & 0.6056 & 0.6248 & 0.3904 & 0.021 & 0.0515 \\
Karate & 0.402 & 0.3547 & 0.3806 & 0.3599 & -0.0186 & 0.0312 \\
Football & 0.3805 & 0.5521 & 0.5497 & 0.3499 & 0.0093 & 0.0022 \\
Dolphin & 0.3805 & 0.4985 & 0.4954 & 0.3847 & 0.0037 & 0.0146 \\
Strike & 0.55 & 0.4861 & 0.5557 & 0.4123 & 0.1147 & 0.0851 \\
Sawmill  & 0.404 & 0.4158 & 0.55 & 0.3869 & 0.0264 & 0.004 \\
}\lookm

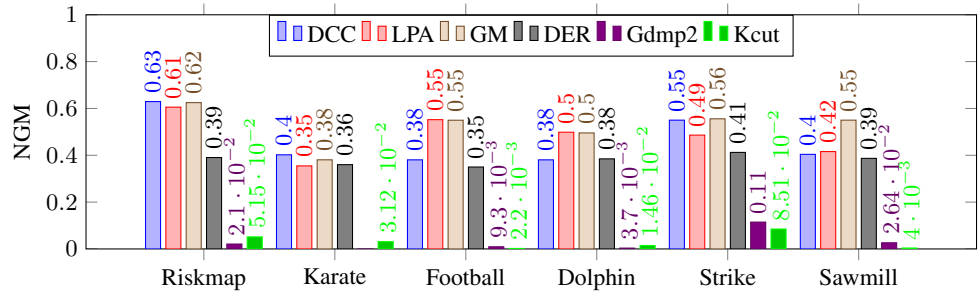
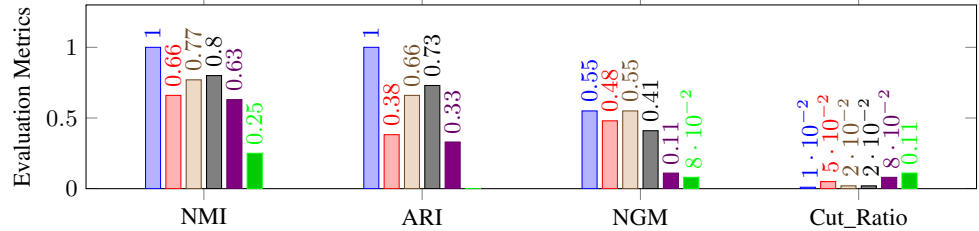
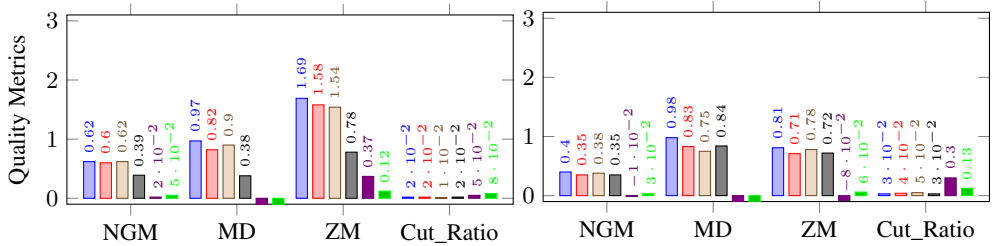
\begin{figure}[!hb]
\centering
\begin{subfigure}{1.1\textwidth}
\centering
\begin{tikzpicture}[remember picture]
    \begin{axis}[
            ybar,
           bar width=.2cm,
            width=\textwidth,
            height=.35\textwidth,
            enlarge x limits=0.18,
            legend style={at={(0.5,1)},
                anchor=north,legend columns=6,legend cell align=left},
            symbolic x coords={Riskmap,Karate,Football,Dolphin, Strike, Sawmill},
            xtick=data,
             x tick label style={rotate=00,anchor=north},
            nodes near coords align={vertical},
            ymin=0,ymax=1.0,
            ylabel={NGM},
            nodes near coords,
            every node near coord/.append style={rotate=90, anchor=west}
        ]
        \addplot table[x=NGM,y=DCC]{\lookm};
        \addplot table[x=NGM,y=LPA]{\lookm};
        \addplot table[x=NGM,y=GM]{\lookm};
        \addplot table[x=NGM,y=DER]{\lookm};
        \addplot table[x=NGM,y=Gdmp2]{\lookm};
        \addplot table[x=NGM,y=Kcut]{\lookm};
        \legend{DCC,LPA,GM,DER,Gdmp2,Kcut}
    \end{axis}
\end{tikzpicture}
\captionsetup{skip=5pt}
\caption{\scriptsize Comparative analysis of community detection algorithms based on Newman Girvan Modularity.} 
\label{nm}
\end{subfigure}%

\pgfplotstableread[row sep=\\,col sep=&]{
Evaluation Metrics & DCC & LPA & GM & DER & Gdmp2 & Kcut\\
NMI & 1 & 0.66 &  0.77 & 0.80 & 0.63 & 0.25\\
ARI & 1 & 0.382 & 0.66 & 0.73 & 0.33 & -0.05 \\
NGM &  0.55 & 0.48 &  0.55 & 0.41 &  0.11 & 0.08\\
Cut\_Ratio & 0.01 &  0.05 & 0.02 & 0.02 & 0.08 &  0.11\\
}\lookm
\begin{subfigure}{1.1\textwidth}
\centering
\begin{tikzpicture}[remember picture]
    \begin{axis}[
            ybar,
           bar width=.2cm,
            width=\textwidth,
            height=.3\textwidth,
            enlarge x limits=0.18,
            symbolic x coords={NMI, ARI, NGM, Cut\_Ratio},
            xtick=data,
             x tick label style={rotate=00,anchor=north},
            nodes near coords align={vertical},
           ymin=0,ymax=1.3,
            ylabel={Evaluation Metrics},
            nodes near coords,
            every node near coord/.append style={rotate=90, anchor=west}
        ]
        \addplot table[x=Evaluation Metrics,y=DCC]{\lookm};
        \addplot table[x=Evaluation Metrics,y=LPA]{\lookm};
        \addplot table[x=Evaluation Metrics,y=GM]{\lookm};
        \addplot table[x=Evaluation Metrics,y=DER]{\lookm};
        \addplot table[x=Evaluation Metrics,y=Gdmp2]{\lookm};
        \addplot table[x=Evaluation Metrics,y=Kcut]{\lookm};
    \end{axis}
\end{tikzpicture}
\captionsetup{skip=5pt}
\caption{\scriptsize Comparative analysis based on quality  and accuracy on  Strike dataset.}
\label{qa}
\end{subfigure}


\pgfplotstableread[row sep=\\,col sep=&]{
Quality Metrics & DCC & LPA & GM & DER & Gdmp2 & Kcut\\
NGM & 0.62 &  0.60 & 0.62 &  0.39 & 0.02 & 0.05\\
MD & 0.97 & 0.82 & 0.90 & 0.38 & -0.15 & -1.34\\
ZM & 1.69 &  1.58 & 1.54 & 0.78 & 0.37 & 0.12 \\
Cut\_Ratio & 0.02 & 0.02 & 0.01 & 0.02 & 0.05 & 0.08 \\
}\lookm
\begin{subfigure}{0.56\textwidth}
\begin{tikzpicture}[remember picture]
    \begin{axis}[
            ybar,
           bar width=.15cm,
            width=1.1\textwidth,
            height=0.6\textwidth,
            enlarge x limits=0.2,
            symbolic x coords={NGM, MD, ZM, Cut\_Ratio},
            xtick=data,
             x tick label style={rotate=00,anchor=north},
            nodes near coords align={vertical},
            ymin=-0.1,ymax=3.1,
            ylabel={Quality Metrics},
            nodes near coords,
            every node near coord/.append style={font=\tiny,rotate=90, anchor=west}
        ]
        \addplot table[x=Quality Metrics,y=DCC]{\lookm};
        \addplot  table[x=Quality Metrics,y=LPA]{\lookm};
        \addplot  table[x=Quality Metrics,y=GM]{\lookm};
        \addplot table[x=Quality Metrics,y=DER]{\lookm};
         \addplot table[x=Quality Metrics,y=Gdmp2]{\lookm};
        \addplot table[x=Quality Metrics,y=Kcut]{\lookm};
    \end{axis}
\end{tikzpicture}

\captionsetup{skip=8pt}
\caption{\scriptsize Comparative analysis based on quality on Riskmap dataset.}
\label{ris}
\end{subfigure}%
\pgfplotstableread[row sep=\\,col sep=&]{
Quality Metrics & DCC & LPA & GM & DER & Gdmp2 & Kcut\\
NGM & 0.40 & 0.35 & 0.38 & 0.35 & -0.01 & 0.03\\
MD & 0.98 & 0.83 & 0.75 & 0.84 & -3.25 & -3.33 \\
ZM & 0.81 & 0.71 & 0.78 & 0.72 & -0.08 & 0.06\\
Cut\_Ratio & 0.03 & 0.04 & 0.05 & 0.03 & 0.30 & 0.125\\
}\lookm
\begin{subfigure}{0.56\textwidth}
\begin{tikzpicture}[remember picture]
    \begin{axis}[
            ybar,
           bar width=.15cm,
            width=1.1\textwidth,
            height=0.6\textwidth,
            enlarge x limits=0.2,
            symbolic x coords={NGM, MD, ZM, Cut\_Ratio},
            xtick=data,
             x tick label style={rotate=00,anchor=north},
            nodes near coords align={vertical},
            ymin=-0.1,ymax=3.1,
            ylabel={},
            nodes near coords,
            every node near coord/.append style={font=\tiny,rotate=90, anchor=west}
        ]
        \addplot table[x=Quality Metrics,y=DCC]{\lookm};
        \addplot table[x=Quality Metrics,y=LPA]{\lookm};
        \addplot table[x=Quality Metrics,y=GM]{\lookm};
        \addplot table[x=Quality Metrics,y=DER]{\lookm};
         \addplot table[x=Quality Metrics,y=Gdmp2]{\lookm};
        \addplot table[x=Quality Metrics,y=Kcut]{\lookm};
    \end{axis}
\end{tikzpicture}
\captionsetup{skip=10pt}
\caption{\scriptsize Comparative analysis based on quality on Karate dataset.}
\label{kar}
\end{subfigure}%
\captionsetup{margin=0.2cm}
\caption{Comparative analysis based on different evaluation metrics on real-world datasets, NGM:~Newman Girvan Modularity, MD:~Modularity Density, ZM:~Z Modularity.} 
\label{figthree}
\end{figure}

\section{Experimental Setup}
\label{four}
In this section, we shall discuss about the experimental setup. Here, experiments are conducted to evaluate the comparative performance of $DCC$ with respect to the baseline community detection algorithms.  Evaluation is carried from three perspectives such as, community detection algorithms, real-world datasets and evaluation metrics. We have selected community detection algorithms that are based on network structure, modularity optimization, random walk and neighborhood information of nodes.

\textbf{Community detection algorithms:} ~Algorithms based on diffusion such as, Label Propagation Algorithm $(LPA)$~\cite{cordasco2010community}; modularity maximization based algorithms such as, Greedy-modularity $(GM)$~\cite{clauset2004finding} and $Kcut$~\cite{ruan2007efficient}; random walk based algorithm such as, Diffusion Entropy Reducer $(DER)$~\cite{kozdoba2015community} and $Gdmp2$~\cite{chen2010dense} based on clustering nodes. These algorithms are selected to analyze and compare the performance of $DCC$ in terms of modularity, neighborhood information of nodes and cascade information. Moreover, the evaluation of communities are carried in two perspectives such as quality and accuracy. Evaluation of community quality is performed in terms of number of internal and external connections. Quality evaluation do not require ground truth information. Whereas, accuracy evaluation requires ground truth information. The following evaluation metrics have been considered for our experimentation purpose.\\

\textbf{Evaluation metrics:}~Quality metrics based on internal connections only such as, NGM, Modularity Density, Z Modularity; external connections based quality metrics such as Cut\_Ratio have been used. Moreover, accuracy metrics such as, Normalized Mutual Information (NMI) and Adjusted Random Index (ARI) have been used~\cite{newman2004finding,miyauchi2016z,fortunato2010community,hubert1985comparing}. Next, the above mentioned community detection algorithms are tested on several real-world datasets such as, riskmap, karate, football, dolphin, strike and sawmill are summarized in Table~\ref{dat}. These datasets are publicly available in online repositories such as SNAP~\cite{snapnets}. The reason to select these datasets is availability of ground-truth information and for ease of performance evaluation by visualization.

\section{Result Analysis}
\label{five}

The comparative results of $DCC$ algorithm with respect to the baseline algorithms considered in this paper have been represented in Fig.~\ref{figthree}. Before discussing about the results obtained by incorporation of several evaluation metrics, let us first try to interpret the result of $DCC$ on karate dataset. $DCC$ gives three set of most densely connected communities on karate dataset.  From this result, we can say that $DCC$ works excellently to identify all groups of densely connected nodes irrespective of the size of such groups.

Let us try to comprehend the results of $DCC$ obtained by incorporation of several evaluation metrics one by one. Firstly, if we consider the result represented in Fig.~\ref{nm}, $DCC$ gives the best Newman Girvan Modularity (NGM) score on riskmap, karate and strike datasets. Whereas, the results on football, dolphin and sawmill is comparative low. The reason for this is that $DCC$ algorithm explicitly identifies the densely connected group of nodes without considering the number of nodes in the corresponding group. Hence, low modularity value does not infer low quality communities. The good performance of $DCC$ is also justified by the results obtained on the remaining quality metrics on football, dolphin and sawmill network.

Consider the results represented in Fig.~\ref{qa}, clearly $DCC$ gives the maximum NMI, ARI, NGM score and minimum Cut\_Ratio score as compared to the baseline algorithms on strike dataset. Therefore, from these results, it is obtained that $DCC$ shows excellent performance in terms of quality and accuracy on strike dataset. Similarly, the results of $DCC$ in terms of these quality and accuracy metrics on other datasets are quite good. Also, results based on different variants of modularity such as; NGM, MD, ZM and results based on Cut\_Ratio on riskmap and karate dataset as shown in Fig.~\ref{ris} and Fig.~\ref{kar} respectively is implication of the excellent performance of $DCC$. Also, though we have not used any modularity based optimization concept in $DCC$ algorithm, but the excellent modularity results is self explanatory of the effectiveness of Neighborhood Similarity measure and Preferential Membership method. 

\section{Conclusion}
\label{six}

In this paper, a novel tie strength guided cascade generation approach for community detection called $DCC$ has been developed. Depending on cascades that are generated, a new method called Preferential Membership has been designed. The interpretation of communities obtained by $DCC$ algorithm assures it's ability to identify densely connected communities irrespective of the size of such communities. We have considered six real-world datasets, five baseline algorithms, four quality evaluation metrics and two accuracy metrics for performance evaluation. The results given by $DCC$ confirms effectiveness of the proposed tie strength measure, cascade generation strategy and preferential membership method. In future, we shall examine the performance of $DCC$ on large real-world networks and synthetic networks and examine it's performance.
\bibliographystyle{unsrt}
\bibliography{bibtex}

\end{document}